\documentclass[prd,aps,tightenlines,superscriptaddress,amsmath,amssymb,amsfonts]{revtex4-2}
\usepackage{amssymb,amsmath,amsthm,amsbsy,epsfig,color,graphicx,times}
\usepackage[ansinew]{inputenc}
\usepackage[english]{babel}
\usepackage{color}
\usepackage{braket}
\usepackage{tabularx}
\usepackage{array}
\usepackage{mhchem}
\usepackage{multirow}

\begin{document}

\title{Neutrino capture on tritium as a probe of flavor vacuum condensate and dark matter}

\author{A. Capolupo}
\email{capolupo@sa.infn.it}
\affiliation{Dipartimento di Fisica ``E.R. Caianiello'' Universit\`{a} di Salerno, and INFN -- Gruppo Collegato di Salerno, Via Giovanni Paolo II, 132, 84084 Fisciano (SA), Italy}

\author{A. Quaranta}
\email{anquaranta@unisa.it}
\affiliation{Dipartimento di Fisica ``E.R. Caianiello'' Universit\`{a} di Salerno, and INFN -- Gruppo Collegato di Salerno, Via Giovanni Paolo II, 132, 84084 Fisciano (SA), Italy}

\begin{abstract}

We show that the study of neutrino capture on tritium, for non-relativistic neutrinos, can allow to distinguish among the various neutrino models, eventually prove the quantum field theory condensation effects and  permit to
test the hypothesis according to which the flavor vacuum  energy  gives a contribution to the dark matter of the universe.
Indeed, we show that the capture rate depends on the neutrino model considered, and that it brings an imprint of the flavor vacuum condensate.
  Experiments like PTOLEMY, designed to reveal  the cosmic neutrino background, then can give an indication of the existence of the dark matter component induced by neutrino mixing. 

\end{abstract}

\maketitle

\section{Introduction}

Neutrino oscillations, which by now have accumulated a significant amount of experimental confirmations \cite{Osc1,Osc2,Osc3,Osc4}, are a compelling evidence for non-vanishing neutrino masses and for physics beyond the standard model of particles \cite{BSM1,BSM2,BSM3,BSM4}. The quantum mechanical theory of neutrino oscillations was originally put forward by Pontecorvo \cite{Pontecorvo1,Pontecorvo2} and is today widely accepted in its $3$-flavor guise. Notwithstanding the consensus on the basic mechanism of neutrino oscillations, neutrino physics is still beset by a number of open issues. The origin of neutrino masses, the fundamental nature (Dirac or Majorana) of neutrinos, the existence of additional flavors, as well as the correct field theoretical description of neutrino mixing, remain debated. On the other hand neutrinos play a major role in cosmology and astrophysics. They figure within leptogenesis scenarios to explain the original baryon asymmetry and sterile neutrinos may contribute to dark matter \cite{CosmoNeutrino1,CosmoNeutrino2}.

Originally introduced to explain the rotation curves of galaxies \cite{CDM2}, dark matter \cite{CDM1,CDM3} is one of the long-standing puzzles in modern cosmology. Dark matter is modelled as a pressure-less perfect fluid and is believed to account for about $85 \%$ of the total matter content of the universe. There is no general consensus on the composition of the mysterious substance. Several explanations for the ``missing matter'' have been proposed, including primordial black holes \cite{PDM1,PDM2}, supersymmetric particles \cite{SSDM}, axions \cite{ADM1,ADM2,ADM3} and sterile neutrinos \cite{CosmoNeutrino2,SNDM}.

In recent years a new connection between neutrino mixing and cosmology has emerged. One of the possible quantum field theory (QFT) models of neutrino mixing, based on the flavor Hilbert space \cite{FF1,FF2,FF3}, involves a condensed vacuum state, the ``flavor vacuum''. Due to its condensate structure the flavor vacuum yields a purely quantum field theoretical contribution to the energy-momentum tensor of matter. Remarkably, the energy-momentum tensor associated to the flavor vacuum is that of a pressure-less perfect fluid \cite{FDM,FDM1,FDM2,FDM3,FDM4}. Consequently the flavor vacuum may possibly represent a dark matter component.

Testing this hypothesis proves to be a formidable task for the present possibilities. The same mechanism underlying the flavor vacuum condensation is responsible for QFT modifications to the neutrino oscillation formulas. Yet these modifications are mostly negligible except for non-relativistc neutrinos $E_{\nu} \simeq m_{\nu}$, whereas neutrino oscillations are usually accessible at much larger neutrino energies $E_{\nu} \gg m_{\nu}$. This rules out oscillation experiments as a possible probe of the condensation mechanism. There are, however, experiments aimed at the revelation of non-relativistic neutrinos, like PTOLEMY \cite{PTOLEMY}, where the QFT effects may be observable.

Here, we discuss the possibility of testing the QFT of neutrino mixing and the related dark-matter-like contribution by means of low energy neutrino experiments such as PTOLEMY. We study the QFT condensation effect on the basic detection process, the neutrino capture on tritium. We show that the capture rate depends on the neutrino model considered, and, in particular, that it bears a trace of the flavor vacuum condensate. We argue that low energy neutrino experiments may not only distinguish among the possible QFT formulations of neutrino mixing, but also indirectly probe the hypothesis of the flavor vacuum as a dark matter component.

The paper is structured as follows. In section II we elucidate the connection between neutrino mixing and dark matter, recalling the main results in this respect. In section III we present the basic amplitude for the inverse beta decay and discuss the four possible choices of neutrino states according to the known neutrino mixing schemes. In section IV we compute the capture cross sections and the capture rates corresponding to each of the four choices of neutrino states, showing the observable effects due to the flavor vacuum condensation. The last section is devoted to the conclusions.

\section{Cosmological effects from the QFT of neutrino mixing}

The quantum mechanical theory of neutrino oscillations pioneered by Pontecorvo \cite{Pontecorvo1,Pontecorvo2} is sufficient to explain most of the observed phenomenology regarding neutrinos. The corresponding quantum field theory, however, is still lacking a universally accepted formulation. In particular, it is debated whether the basic quantum fields are the massive fields $\nu_1 , \nu_2 , \nu_3$ which propagate freely in vacuum, or the flavor fields $\nu_e , \nu_{\mu} , \nu_{\tau}$, which are those appearing in the weak interaction vertices. In the ``flavor fock space'' approach \cite{FF1,FF2,FF3} flavor fields are the fundamental entities and the usual mixing relations between neutrino states are elevated to relations between the quantum fields:

\begin{equation}\label{FieldMixing}
 \nu_e (x) = \cos \Theta \ \nu_{1} (x) + \sin \Theta \ \nu_2 (x) \ ; \ \ \ \ \   \nu_{\mu} (x) = \cos \Theta \ \nu_{2} (x) - \sin \Theta \ \nu_1 (x) \ ,
\end{equation}
where we have considered two flavors with mixing angle $\Theta$ for definiteness. Because mixing is here stated at the level of fields, the corresponding $1$-particle states are mixed everywhere on spacetime, that is, regardless of the particular spacetime position $x$. In the flavor Fock space approach,  Eq.\eqref{FieldMixing} is recast in terms of a mixing generator $G_{\Theta} (t)$ as \cite{FF1,FF2,FF3} 

\begin{equation*}
\nu_e(x) = G_{\Theta}^{-1} (t) \nu_{1} (x) G_{\Theta} (t) \  ; \ \ \ \ \nu_{\mu}(x) = G_{\Theta}^{-1} (t) \nu_{2} (x) G_{\Theta} (t)
\end{equation*}
whose form is found to be $G_{\Theta} (t) = \exp \left[\Theta \left( \left(\nu_2,\nu_1\right)_t - \left(\nu_1,\nu_2\right)_t\right)\right]$, with $(,)_t$ denoting the Dirac inner product on the hypersurface $x^0 = t$. The action of the generator is then taken to \emph{define} the flavor annihilators as
\begin{eqnarray*}
 a_{\pmb{k},\nu_e}^{s} (t) &=& G_{\Theta}^{-1} (t) a_{\pmb{k},1}^s G_{\Theta}(t) = \cos \Theta \ a_{\pmb{k},1}^s + \sin \Theta \left(U_{\pmb{k}}^* (t) a_{\pmb{k},2}^s + \epsilon^{s} V_{\pmb{k}}(t)b^{s \dagger}_{-\pmb{k},2}\right)
\end{eqnarray*}
and similar for $a_{\pmb{k},\nu_{\mu}}^s, b_{\pmb{k},\nu_{e}}^s, b_{\pmb{k},\nu_{\mu}}^s$, with $s= \pm 1/2$ and $\epsilon^s = (-1)^{s-\frac{1}{2}} $. The flavor operators are then expressed in terms of the mass operators through a Bogoliubov transformation nested into a rotation. The terminology is justified by noting that the coefficients $U_{\pmb{k}}(t) = (u_{\pmb{k},2}, u_{\pmb{k},1})_t, V_{\pmb{k}}(t) = (u_{\pmb{k},1}, v_{\pmb{k},2})_t$, given by the scalar products of modes with positive $u$ and negative energy $v$, satisfy the identity $|U_{\pmb{k}} (t)|^2 + |V_{\pmb{k}} (t)|^2 = 1$.
The appearance of a mass creation operator in the expression of the flavor annihilators signals that the representation defined by the flavor operators is unitarily inequivalent to the representation defined by the mass annihilators. In particular, the flavor annihilators define a time-dependent vacuum state, the flavor vacuum $\ket{0_F (t)}$, which is distinct from the mass vacuum $\ket{0_M}$.
Remarkably, the flavor vacuum has the structure of a condensate of particle-antiparticle pairs with definite masses \cite{FF1,FF2,FF3}. This is apparent if one computes the number expectation values on the flavor vacuum
\begin{equation}\label{cond}
N^F_{\pmb{k},i}  = \bra{0_F (t)} a_{\pmb{k},i}^{\dagger} a_{\pmb{k},i} \ket{0_F(t)} = \sin^2 \Theta |V_{\pmb{k}}|^2 = \bra{0_F (t)} b_{\pmb{k},i}^{\dagger} b_{\pmb{k},i} \ket{0_F(t)} \ ,
\end{equation}
which holds for any $i=1,2$ and any $\pmb{k}$. The quantity $\sin^2 \Theta |V_{\pmb{k}}|^2$ plays the role of a condensation density.
Notice that $N^F_{\pmb{k},i} $ can be equivalently expressed as $N^F_{\pmb{k},i}  = \bra{0_{M}  } a_{\pmb{k},\sigma}^{\dagger} a_{\pmb{k},\sigma} \ket{0_{M}}$, with $\sigma =e,\mu$. This simply follows from equation \eqref{cond} by inserting the identity $1 = G_{\Theta} (t) G_{\Theta}^{-1} (t)$ on the left and on the right of the number operator $a_{\pmb{k},i}^{\dagger} a_{\pmb{k},i}$. 

It can be shown that, as a consequence of the condensate structure, a non-vanishing energy momentum tensor $T_{\mu \nu}$ is associated to the flavor vacuum. Specifically, the pressure $p$ is found to vanish, due to a vanishing expectation value of the spatial components of the energy momentum tensor $\bra{0_F (t)} T^{i}_{i} \ket{0_F (t)} = 0$. On the other hand, the flavor vacuum is associated with a non zero energy density given by \cite{FDM1,FDM2,FDM3,FDM4}
\begin{equation}\label{Flatenergydensity}
 \rho = \bra{0_F (t)} T^{0}_{0} \ket{0_F (t)} = 8 \sin^2 \Theta \int d^3 k \left(\omega_{\pmb{k},1}+ \omega_{\pmb{k},1}\right) |V_{\pmb{k}}|^2 \ .
\end{equation}
Here $\omega_{\pmb{k},j} = \sqrt{\pmb{k}^2 + m_j^2}$ for $j=1,2$. As it is evident from equation \eqref{Flatenergydensity}, the energy density vanishes in absence of mixing $\Theta = 0$ and for $V_{\pmb{k}} = 0$. Therefore it is a direct consequence of the unitarily inequivalence between the flavor and the mass representations. Not only is the flavor vacuum associated with a non vanishing energy momentum tensor, but its components satisfy the equation of state \cite{FDM1,FDM2,FDM3,FDM4}
\begin{equation}\label{FlatEquationofState}
 p = w \rho = 0 \ \ \longrightarrow \ \ w = 0   \ \ \ (\rho \neq 0) \ ,
\end{equation}
which is typical of dust and cold dark matter \cite{CDM1,CDM2,CDM3}. The results \eqref{Flatenergydensity} and \eqref{FlatEquationofState} were derived in flat space, but already provide an indication of the possibility that the flavor vacuum may contribute to cold dark matter, due to its condensate structure.

A stronger hint to such a possibility comes from the curved spacetime generalization of the flavor fock space approach \cite{FDM,Curved1}. Much of the formalism, including the condensate structure of the flavor vacuum, is common to the flat space theory. The main advantage is that in curved spacetime the energy momentum tensor associated to the flavor vacuum can be consistently studied within the semiclassical approach, where it appears on the right hand side of the Einstein field equations as a proper source term. Even if the analysis, in this respect, is still ongoing and must be extended to other metrics, recent results show that also in curved space the flavor vacuum satisfies the equation of state of cold dark matter.

For a spatially flat Friedmann-Robertson-Walker metric with De Sitter evolution, at late times, the energy density takes the form \cite{FDM}

\begin{eqnarray}\label{EtaEtaFunction4}
  \nonumber \rho (\tau) &\simeq& i \sin^2 \Theta \sum_{\lambda} \int d^3 p |\Xi_p(\tau_0)|^2 \left(i \frac{H_0^3 \tau^3}{2 \pi^3} \right) \sum_{j=1,2} m_j \tanh \left(\frac{\pi m_j}{H_0} \right) \\
 \nonumber &+& \frac{i}{2} \sin^2 \Theta \sum_{\lambda} \int d^3 p \left[\Xi^*_p (\tau_0) \Lambda_p(\tau_0) \left(\frac{-im_1 H_0^3 \tau^3}{2 \pi^3 \cosh \left( \frac{\pi m_1}{H_0} \right)} \right) - c.c. \right] \\ \nonumber
  &-& \frac{i}{2} \sin^2 \Theta \sum_{\lambda} \int d^3 p \left[\Xi^*_p (\tau_0) \Lambda^*_p(\tau_0) \left(\frac{-im_2 H_0^3 \tau^3}{2 \pi^3 \cosh \left( \frac{\pi m_2}{H_0} \right)} \right) - c.c. \right] \  .
\end{eqnarray}
where $\tau$ is the conformal time, $H_0$ is the constant Hubble factor and $\Lambda, \Xi$ are the curved space counterparts of $U, V$ (Bogoliubov coefficients). For the details, we refer the reader to the work \cite{FDM}. The essential point of Eq.\eqref{EtaEtaFunction4} is the dependence of the energy density on the mixing angle and on the Bogoliubov coefficients. Just as in flat space, $\rho = 0$ whenever $\Theta = 0$, or $\Xi = 0$. In addition, the pressure $p$ is found to vanish, so that, also in this case, the flavor vacuum satisfies the equation of state \eqref{FlatEquationofState}, typical of cold dark matter. Ongoing studies are aimed at proving the same relation in other classes metrics.

The results obtained both in flat and in curved spacetime show that the condensate associated to the flavor vacuum might play a role as a cold dark matter component. Given the cosmological and astrophysical relevance of such a possibility, an experimental test of the underlying theory is desirable. Besides the energy-momentum considerations related to the flavor vacuum, the flavor Fock space approach predicts modified oscillation formulas, both in flat and in curved space. These modifications are hard to observe directly, because their magnitude is negligible at almost all the energy scales, except when the neutrino energy is close to the neutrino masses $E \sim m_j$. On the other hand observed neutrinos are all extremely relativistic $E \gg m_j$, so that a direct test of the theory through oscillation experiments is unviable, at least for the current possibilities. A direct test of the cosmological implications is likewise hard to attain due to the large scales involved. For these reasons previous attempts at probing the flavor Fock space model have focussed on analogous systems. In \cite{Rydberg} it was proposed that Rydberg atoms be used to simulate neutrino mixing and reproduce the resulting condensate, which would leave a signature on the thermodynamic quantities of the system. Recently the impact of the flavor condensate on the beta decay spectrum has also been explored \cite{Lee2020}.

In this paper we propose that the neutrino capture on tritium be used to probe the flavor Fock space model. Due to the sensitivity of the process to extremely low energy neutrinos, there is a hope that the quantum field theoretic corrections, mostly relevant for non-relativistic ($E \sim m_j$) neutrinos may show up in capture experiments. An experiment of this kind is PTOLEMY \cite{PTOLEMY} which is aimed at revealing the extremely non-relativistic neutrinos of the cosmic neutrino background (C$\nu$B). In the next sections we show that, if the flavor Fock space approach is valid, the quantum field theoretic corrections show up in the capture rate for neutrinos, which assumes different values according to the neutrino model considered. Experiments aimed at the revelation of the C$\nu$B  can therefore distinguish among the various neutrino models and eventually validate the hypothesis of the flavor vacuum as a dark matter component.

\section{Hamiltonian and neutrino states}

We shall be primarily interested in non-relativistic (low energy) neutrinos which constitute the C$\nu$B \cite{CNB1,CNB2} and may be probed by the PTOLEMY experiment \cite{PTOLEMY}. The basic reaction for neutrino capture is the inverse beta decay $\nu + n \rightarrow e^{-} + p$. Due to the low neutrino energies involved, the reaction can be safely described with the current-current interaction Hamiltonian \cite{Long2014}:
\begin{equation}\label{Interaction Hamiltonian}
 H_I = \frac{G_F}{\sqrt{2}} V_{ud} \left[\bar{p}(x) \gamma_{\mu} \left( f(0) -g(0)\gamma^5 \right)n(x) \right] \left[\bar{e} (x) \gamma^{\mu} \left(1-\gamma^5\right)\nu_e (x) \right] + h.c. \ .
\end{equation}
Here $G_F = 1.166 \times 10^{-5} \mathrm{GeV}^{-2}$ is the Fermi constant, $V_{ud}$ is the CKM matrix element, $f(q)$ and $g(q)$ are nuclear form factors \cite{PDG} and $n(x), p(x), e(x), \nu_e (x)$ are respectively the neutron, the proton, the electron and the neutrino field. We shall work in the interaction picture. The tree level amplitude $\mathcal{A}$ is usually defined as
\begin{equation}\label{Interaction Amplitude}
 S_{FI} = I_{FI} + (2\pi)^4 i \delta^4 (P_F^{\mu} - P_I^{\mu}) \mathcal{A}_{FI} = I_{FI} - i \left( \int d^4 x H_I (x) \right)_{FI}
\end{equation}
where the indices $FI$ denote the matrix elements between the initial $I$ and final state $F$, $S$ is the S-matrix, $I$ is the identity matrix and the Dirac delta enforces total $4$-momentum conservation. For reasons that will become apparent in a moment, we work directly with the unnormalized matrix element $S_{FI} = i \left( \int d^4 x H_I (x) \right)_{FI}$. Assuming, for simplicity, two neutrino flavors, the electron neutrino field can be written in terms of the mass neutrino fields as $\nu_e(x) = \cos \Theta \ \nu_1 (x) + \sin  \Theta \ \nu_2 (x)$, with mixing angle $\Theta$. The initial and final states for the inverse beta decay are
\begin{equation}\label{Interaction States}
 \ket{I} = \ket{0_e} \ket{0_p} \ket{n_{P_n, s_n}} \ket{\nu_{e ; P_{\nu}, s_{\nu}}(I)} \ , \ \ \ \ \ \ket{F} = \ket{e_{P_e, s_e}} \ket{p_{P_{p},s_p}} \ket{0_n} \ket{0_\nu (F)}
\end{equation}
where the $P_j$ are momentum indices, the $s_j$ spin indices for $j=e,n,p,\nu$ and the tensor product is understood. The arguments $I, F$ in the neutrino states are inserted to keep track of the two asymptotic limits $t_{I,F} \rightarrow \mp \infty$.

The difference in the various approaches to neutrino mixing is essentially in the definition of the neutrino flavor states. We distinguish four cases:

\begin{enumerate}
 \item \emph{Decoupled Pontecorvo states}

 This is the case considered in the Ref.\cite{Long2014}. The neutrinos are originally produced as Pontecorvo flavor states $\ket{\nu_e} = \cos \Theta \ket{\nu_1} + \sin \Theta \ket{\nu_2}$ but are quickly decoupled due to the different propagation velocities for $\nu_1$ and $\nu_2$. At the observation time, these states have completely decohered into mass eigenstates, and the inverse beta decay takes the form, $\nu_j + n \rightarrow e^{-} + p$. For the effects of decoherence on neutrino oscillations see also ref. \cite{Deco}. The only effect of mixing is in the interaction Hamiltonian \eqref{Interaction Hamiltonian}, where the mixing matrix determines the fractions of $\nu_1$ and $\nu_2$ that interact as electron neutrinos. It is important to remark that in this case the neutrinos are mixed only at the level of states.

 \item \emph{Pontecorvo states}

 In this case neutrinos are not only produced, but also interact as Pontecorvo flavor states. The mixing is to be understood at the level of fields $\nu_e (x) = \cos \Theta \ \nu_1 (x) + \sin \Theta \ \nu_2 (x)$ and is therefore ``eternal'' (no decoherence).  The neutrino states entering the amplitude \eqref{Interaction Amplitude} are of the form $\ket{\nu_e} = \cos \Theta \ket{\nu_1} + \sin \Theta \ket{\nu_2}$. Such a definition neglects the spinorial nature of the neutrino states and amounts to the action of the flavor creation operator $a_e^{\dagger} = \cos \Theta \ a_1^{\dagger} + \sin \Theta \ a_2^{\dagger}$ on the neutrino vacuum state $\ket{0_{\nu}}$.

 \item \emph{Pontecorvo-Dirac states}

 Also in this case mixing is at the level of fields. Here one does not neglect the spinorial nature of the neutrino states and sets $a_e^{\dagger}= (u_1, \nu_e (x))^{\dagger} \sim \cos \Theta a^{\dagger}_1 + \sin \Theta \  U a^{\dagger}_2 + \sin \Theta \ V b_2$, with $(,)$ denoting the Dirac inner product and $b_2$ the anti-$\nu_2$ destruction operator. The coefficients $U$ and $V$ are the scalar products $U \sim (u_1 , u_2), V \sim (u_1, v_2)$. When acting on the neutrino vacuum, the creation operator $a^{\dagger}_e$ produces the state $\ket{\nu_e} = \cos \Theta \ \ket{\nu_1} + \sin \Theta \ U \ket{\nu_2}$. This assumes a unique neutrino vacuum state, the mass vacuum $\ket{0_{\nu}} \equiv \ket{0_M}$ annihilated by $a_1, a_2, b_1, b_2$.
 Notice that in this case, since the flavor vacuum and the mass vacuum coincide $\ket{0_F(t)} = \ket{0_M}$, then
  \begin{equation}
N^F_{\pmb{k},i}  = \bra{0_{M}} a_{\pmb{k},i}^{\dagger} a_{\pmb{k},i} \ket{0_{M}} = 0 \ ,
\end{equation}
and the corresponding vacuum energy is equal to zero.
One can also compute the expectation value of the operator $a_{\pmb{k},\sigma}^{\dagger} a_{\pmb{k},\sigma}$, with $\sigma =e,\mu$, on the vacuum state and to obtain, $ \Upsilon^F_{\pmb{k},i} = \bra{0_{M}  } a_{\pmb{k},\sigma}^{\dagger} a_{\pmb{k},\sigma} \ket{0_{M}}=\sin^2 \Theta |V_{\pmb{k}}|^2$.  This result analytically coincides with Eq.(\ref{cond}). However, the physical interpretation of $ \Upsilon^F_{\pmb{k},i} $ is completely different from that of the vacuum condensate density given in Eq.(\ref{cond}). Indeed,  $\Upsilon^F_{\pmb{k},i}$ is not related to the energy momentum tensor for free massive fields and its contribution to the energy of the universe is equal to zero. The vacuum energy (\ref{Flatenergydensity}), (and the corresponding in curved space (\ref{EtaEtaFunction4})),  is due only to the unitarily inequivalence of the mass and flavor representation of the Fock spaces in the infinite volume limit, that for Pontecorvo-Dirac states is absent, since the flavor Hilbert space $H_f$ coincides with the Hilbert space for massive field $  H_m$.

 \item \emph{Flavor Fock space states}

 The mixing is at the level of fields and no decoherence occurs. The flavor creation operator is again given by $a_e^{\dagger} \sim \cos \Theta a^{\dagger}_1 + \sin \Theta \  U a^{\dagger}_2 + \sin \Theta \ V b_2$, but in this case the flavor state is defined by its application on the \emph{flavor vacuum} $\ket{0_{\nu,f}}$ which is distinct from the mass vacuum $\ket{0_{\nu}}$. The flavor vacuum $\ket{0_{\nu,f}(t)}$ depends explicitly on time $t$, whence the necessity of the $I,F$ arguments in Eq. \eqref{Interaction States}.
\end{enumerate}

Only the last case features a condensed flavor vacuum, so that only in this case one has a dark matter contribution from the vacuum of the theory. Mathematically speaking the last case is the most general, as all the others can be obtained by taking the appropriate limits. To exhibit the relevant limits we distinguish the mixing matrix elements in the field $\nu_e = \cos \Theta \nu_1 + \sin \Theta \nu_2$ from those appearing in the definition of the states $\ket{\nu_e} \sim \cos \bar{\Theta} \ket{\nu_1} + \sin \bar{\Theta} \ket{\nu_2}$. When both are present, of course, $\Theta = \bar{\Theta}$. The four cases can then be classified according to the table (I). The last column is there referred to the condensation density $N^F_{\pmb{k},i}$ and the energy density $\rho$ associated to the flavor vacuum.

\begin{table}[h]
\label{TableOne}
\begin{tabular}{|p{3cm}|p{2cm}|p{2cm}|p{1cm}|p{1cm}|p{1cm}|}
\hline
\vspace{0.001mm} Case & \vspace{0.001mm}  $\cos\Theta / \sin \Theta$ & \vspace{0.001mm} $\cos \bar{\Theta} / \sin \bar{\Theta}$ & \vspace{0.001mm} $U$ & \vspace{0.001mm} $V$ & \vspace{0.001mm} $N^F, \  \rho$ \\ [0.5ex]
 \hline
 Decoupled Pontecorvo & $\cos\Theta / \sin \Theta$ & $\rightarrow 1 / 1$  & 1 & 0 & 0\\
 \hline
 Pontecorvo & $\cos\Theta / \sin \Theta$ & $ \cos\Theta / \sin \Theta$ & 1 & 0 & 0\\
 \hline
 Pontecorvo-Dirac & $\cos\Theta / \sin \Theta$ & $ \cos\Theta / \sin \Theta$ & $ \neq 1$ & $\neq 0$ & 0\\
 \hline
 Flavor Fock space & $\cos\Theta / \sin \Theta$ & $ \cos\Theta / \sin \Theta$ & $\neq 1$ & $ \neq 0$ & $ \neq 0$\\
 \hline
\end{tabular}
\caption[Table 1]{Summary of the mixing schemes}
\end{table}

Since all the mixing schemes descend from the flavor Fock space approach in the appropriate limits, we compute the amplitude of eq. \eqref{Interaction Amplitude} and the corresponding cross section in this approach.

\section{Amplitude and cross section}

The object of interest is the unnormalized amplitude

\begin{equation}\label{Amplitude1}
 S_{FI} (T) = - i \left( \int_{-\frac{T}{2}}^{\frac{T}{2}} d x^0 \int d^3 x H_I (x) \right)_{FI}
\end{equation}
where we have split off the time and space integrals. The integration boundary $T$ is to be pushed to $\infty$, but we keep it finite, for the moment, for the sake of clarity. In general, we cannot extract a single $\delta^4$ term from Eq.\eqref{Amplitude1} due to the presence of terms with different energies. Using the generic states of Eq.\eqref{Interaction States} we can write
\begin{eqnarray}\label{Amplitude2}
 \nonumber S_{FI}(T) &=& - i \frac{G_F}{\sqrt{2}} \int_{-\frac{T}{2}}^{\frac{T}{2}} d x^0 \int d^3 x \ V_{ud} \ \eta_{\alpha \beta} \bra{p_{P_p,s_p}} \bar{p}(x) \ket{0_p} \left[\gamma^{\alpha} \left( f - g\gamma^5\right)\right] \bra{0_n} n(x) \ket{n_{P_n,s_n}} \\
 &\times& \bra{e_{P_e,s_e}} \bar{e}(x) \ket{0_e} \left[ \gamma^{\beta} \left(1-\gamma^5 \right) \right] \bra{0_{\nu,f}(t_F)} \nu_e (x) \ket{\nu_{e; P_{\nu}, s_{\nu}}(t_I)} \ .
\end{eqnarray}
The matrix elements of the neutron, the proton and the electron states are computed straightforwardly. As anticipated, the neutrino matrix elements involves two (in principle distinct) time arguments $t_I, t_F$ for the asymptotic states, due to the intrinsic time dependence of the flavor vacuum. The neutrino field is expanded as
\begin{equation}
 \nu_e (x) = \sum_{r_{\nu}} \int \frac{d^3 k}{\sqrt{(2 \pi^3)}} \left[ u^{r_{\nu}}_{\pmb{k}_{\nu};1} a^{r_{\nu}}_{\pmb{k}_{\nu};e} (x^0) + v^{r_{\nu}}_{-\pmb{k}_{\nu};1} b^{r_{\nu} \dagger}_{-\pmb{k}_{\nu};e} (x^0)  \right] e^{i \pmb{k}_{\nu} \cdot \pmb{x}}
\end{equation}
where the spinors are normalized to $u^{\dagger}_j u_j = 2 E_{\nu;j} = 2 \sqrt{\pmb{k}_{\nu}^2 + m_j^2}$ and boldface letters are used to denote spatial 3-vectors. By definition the flavor operators are
\begin{eqnarray*}
 a^{s}_{\pmb{k}, \nu_e} (x^0) &=& e^{- i E_{\nu;1}x^0} \left[\cos \Theta a^{s}_{\pmb{k},1} + \sin \Theta \left(U_{\pmb{k}}^* (x^0) a^{s}_{\pmb{k},2} + \epsilon^{s} V_{\pmb{k}}(x^0) b^{s \dagger}_{-\pmb{k},2} \right) \right] \\
  b^{s}_{-\pmb{k}, \nu_e} (x^0) &=& e^{- i E_{\nu;1}x^0} \left[\cos \Theta b^{s}_{-\pmb{k},1} + \sin \Theta \left(U_{\pmb{k}}^* (x^0) b^{s}_{-\pmb{k},2} - \epsilon^{s} V_{\pmb{k}}(x^0) a^{s \dagger}_{\pmb{k},2} \right) \right]
\end{eqnarray*}
where we have assumed, without loss of generality, the neutrino momentum $\pmb{k}_{\nu}$ along the third axis and $\epsilon^s = (-1)^{s-\frac{1}{2}}$ for spin projections $s= \pm \frac{1}{2}$. The Bogoliubov coefficients are $U_{\pmb{k}}(x^0) = |U_{\pmb{k}}| e^{i \left(E_{\nu; 2} - E_{\nu; 1} \right)x^0}$, $V_{\pmb{k}}(x^0) = |V_{\pmb{k}}| e^{i \left(E_{\nu; 2} + E_{\nu; 1} \right)x^0}$ with $|U_{\pmb{k}}|, |V_{\pmb{k}}|$ given in the references \cite{FF1,FF2,FF3}.

As it stands, with distinct time arguments $t_I, t_F$, the neutrino matrix element cannot be computed straightforwardly. But most importantly such a time difference would take into account the intrinsic oscillation properties of the flavor Fock space states (including the vacuum), which is not appropriate for the evaluation of a scattering amplitude. The ``asymptotic'' neutrino states should not involve oscillations at times before and after the interaction process. In addition, due to the small timescales of the weak interaction, it is reasonable to consider the approximation $t_F \simeq t_I$. We then set $t_F = t_I$ and assume $t_F = t_I = 0$. The neutrino matrix element can now be computed easily (omitting the $t_I / t_F$ time arguments)
\begin{eqnarray}
 \nonumber && \bra{0_{\nu, f}} \nu_e (x) \ket{\nu_{e; P_{\nu}, s_{\nu} }} = \\
 \nonumber && \sum_{r_{\nu}} \int \frac{d^3 k_{\nu}}{\sqrt{2 \pi^3}} e^{i \pmb{k}_{\nu} \cdot \pmb{x}} \left[u^{r_{\nu}}_{\pmb{k}_{\nu};1} \bra{0_{\nu, f}} \left \lbrace a^{r_{\nu}}_{\pmb{k}_{\nu}; \nu_e} (x^0) , a^{s_{\nu} \dagger}_{\pmb{p}_{\nu}; \nu_e}  \right \rbrace \ket{0_{\nu,f}} + v^{r_{\nu}}_{-\pmb{k}_{\nu};1} \bra{0_{\nu, f}} \left \lbrace b^{r_{\nu} \dagger}_{-\pmb{k}_{\nu}; \nu_e} (x^0) , a^{s_{\nu} \dagger}_{\pmb{p}_{\nu}; \nu_e}  \right \rbrace \ket{0_{\nu,f}} \right] = \\
 \nonumber && \frac{e^{i \pmb{p}_{\nu} \cdot \pmb{x}}}{\sqrt{2 \pi^3}} \Bigg[u^{s_{\nu}}_{\pmb{p}_{\nu};1}  \left(\cos \Theta \cos \bar{\Theta} e^{-i E_{\nu;1}x^0} + \sin \Theta \sin \bar{\Theta} \left(|U_{\pmb{p}_{\nu}}|^2 e^{-i E_{\nu;2}x^0} + |V_{\pmb{p}_{\nu}}|^2 e^{i E_{\nu;2}x^0}\right) \right) \\
 \nonumber && + v^{s_{\nu}}_{-\pmb{p}_{\nu};1} \epsilon^{s_{\nu}} \sin \Theta \sin \bar{\Theta} |U_{\pmb{p}_{\nu}}| |V_{\pmb{p}_{\nu}}|\left( e^{-i E_{\nu;2}x^0} - e^{i E_{\nu;2}x^0}\right)\Bigg]
\end{eqnarray}
where we have kept the notation $\Theta, \bar{\Theta}$ to distinguish the mixing angles in the field and in the states. Inserting the neutrino matrix element in Eq.\eqref{Amplitude2} and taking the infinite time limit $T \rightarrow \infty$ we find
\begin{equation}\label{Amplitude3}
 S_{FI} = \delta^4(p_{n}^{\mu} + p_{\nu,1}^{\mu} - p_{p}^{\mu}-p_{e}^{\mu}) M_{\nu_1} + \delta^4(p_{n}^{\mu} + p_{\nu,2}^{\mu} - p_{p}^{\mu}-p_{e}^{\mu}) M_{\nu_2} + \delta^4(p_{n}^{\mu} + \bar{p}_{\nu,2}^{\mu} - p_{p}^{\mu}-p_{e}^{\mu}) M_{\bar{\nu}_2} \ ,
\end{equation}
where $p_{\nu,j}^{\mu} \equiv (E_{\nu;j}, \pmb{p}_{\nu})$, $\bar{p}_{\nu,j}^{\mu} \equiv (-E_{\nu;j}, \pmb{p}_{\nu})$ and
\begin{eqnarray*}
 && M_{\nu_1} = - \frac{G_F}{\sqrt{2}} V_{ud} \eta_{\alpha \beta} \cos \Theta \cos \bar{\Theta} R_{1}^{\alpha \beta} \ ; \ \  M_{\nu_2} = - \frac{G_F}{\sqrt{2}} V_{ud} \eta_{\alpha \beta} \sin \Theta \sin \bar{\Theta} R_{2}^{\alpha \beta} \ ; \ \  M_{\bar{\nu}_2} = - \frac{G_F}{\sqrt{2}} V_{ud} \eta_{\alpha \beta} \sin \Theta \sin \bar{\Theta} \epsilon^{s_{\nu}} L_{2}^{\alpha \beta} \\
 && R^{\alpha \beta}_{j} = \bar{u}^{s_p}_{\pmb{p}_p} \gamma^{\alpha} (f-g\gamma^5)u^{s_n}_{\pmb{p}_n} \bar{u}^{s_e}_{\pmb{p}_e} \gamma^{\beta} (1-\gamma^5) u^{s_{\nu}}_{\pmb{p}_{\nu},j} \ \ \ \ \ \ \ \ \ \ \ \ \ \ \ \ \ \ \ \ \ \ \ \ \ \ \ \ \ \ \ \ \ \ \ \ \ \  L^{\alpha \beta}_{j} = \bar{u}^{s_p}_{\pmb{p}_p} \gamma^{\alpha} (f-g\gamma^5)u^{s_n}_{\pmb{p}_n} \bar{u}^{s_e}_{\pmb{p}_e} \gamma^{\beta} (1-\gamma^5) v^{s_{\nu}}_{-\pmb{p}_{\nu},j} \ .
\end{eqnarray*}
The amplitude of Eq. \eqref{Amplitude3} has the form of Eq. \eqref{appendix1} in the infinite volume and infinite time limits. The cross section is then defined according to Eq. \eqref{appendix2}. In computing the squared amplitudes $|M|^2$, we specialize to the neutron rest frame and neglect the proton recoil \cite{Long2014}, so that the $4$-momenta are $p_{n}^{\mu} \equiv (m_n, 0), p_{p}^{\mu} \simeq (m_p,0), p_e^{\mu} \equiv (E_e , \pmb{p}_e), p_{\nu,j}^{\mu} \equiv (E_{\nu,j}, \pmb{p}_{\nu})$. We also set $\pmb{p}_e \cdot \pmb{p}_{\nu} = p_e p_{\nu} \cos \theta$, sum over the neutron and proton spins and average over the electron spin, in order to get the unpolarized cross section. Then
\begin{eqnarray}\label{Amplitude4}
\nonumber|M_{\nu_1}|^2 &=& 8 G_F^2 |V_{ud}|^2 \cos^2 \Theta \cos^2 \bar{\Theta} m_p m_n E_e E_{\nu,1} \left[(1-2s_{\nu}v_{\nu})(3g^2 + f^2) + (v_{\nu}-2s_{\nu})(f^2 -g^2)v_{e} \cos \theta \right] \\
\nonumber |M_{\nu_2}|^2 &=&  8 G_F^2 |V_{ud}|^2 \sin^2 \Theta \sin^2 \bar{\Theta} |U_{\pmb{p}_{\nu}}|^2 m_p m_n E_e E_{\nu,2} \left[(1-2s_{\nu}v_{\nu})(3g^2 + f^2) + (v_{\nu}-2s_{\nu})(f^2 -g^2)v_{e} \cos \theta \right] \\
|M_{\bar{\nu}_2}|^2 &=& 8 G_F^2 |V_{ud}|^2 \sin^2 \Theta \sin^2 \bar{\Theta} |V_{\pmb{p}_{\nu}}|^2 m_p m_n E_e E_{\nu,2} \left[(1-2s_{\nu}v_{\nu})(3g^2 + f^2) + (v_{\nu}-2s_{\nu})(f^2 -g^2)v_{e} \cos \theta \right]
\end{eqnarray}
where we have introduced the velocities $v_j = \frac{p_j}{E_j}$. The first and the second of the squared amplitudes of Eq. \eqref{Amplitude4} agree with those obtained in \cite{Long2014} except for the factors in $\bar{\Theta}$ and $|U_{\pmb{p}_{\nu}}|$. The interpretation of the last term in Eqs. \eqref{Amplitude3} and \eqref{Amplitude4} requires more care.
The second delta function in Eq. \eqref{Amplitude3} forces the energy conservation as $E_{\nu,2} = E_p + E_e - E_n (> 0)$, but the third delta function involves the opposite sign of the neutrino energy, enforcing $E_{\nu,2} = -(E_p + E_e - E_n)$. Assuming the same electron, proton and neutron energies, the last delta implies a negative energy $E_{\nu,2} = - |E_{\nu,2}| (<0)$.

We acknowledge two possibilities for the third term:
\begin{itemize}
 \item If strictly interpreted as the neutrino energy, this term must be discarded, since by definition only positive energies are possible. This is also the route followed in the Ref. \cite{Lee2020} in computing the beta decay spectrum. Discarding the last term is equivalent to set $V_{\pmb{p}_{\nu}} = 0$, so that one obtains the same result as for the Pontecorvo-Dirac states (see Table I).

 \item If interpreted within the context of the condensate structure of the flavor states, the negative energy $-|E_{\nu,2}|$ may be viewed as the energy associated with a $\nu_2$ ``hole'' in the condensate. This is the kind of reasoning invoked in \cite{WDecay}. The last term is then retained with a change of sign in the energy $E_{\nu,2}$ with respect to the second term. Assuming a negative energy for the third term brings both the second and third delta functions to the form $\delta (E_{n} + |E_{\nu,2}| - E_e - E_p)$.
 It is worth noting that the third term $\delta(E_n - E_{\nu,2}-E_p - E_e) \delta^3(\pmb{p}_{n}+\pmb{p}_{\nu}-\pmb{p}_p - \pmb{p}_e)|M_{\bar{\nu}_2}|^2$ also admits another interpretation. Both the delta function and the amplitude (apart from a $\sin^2 \bar{\Theta} |V_{\pmb{p}_{\nu}}|$ factor) may be considered to describe the decay $n \rightarrow e^{-} + p + \bar{\nu}_2$ for an antineutrino $\bar{\nu}_2$ with (positive) energy $E_{\nu,2}$ and momentum $-\pmb{p}_{\nu}$. If interpreted in this way, the amplitude of Eq. \eqref{Amplitude3} is a (weighed) superposition of the amplitudes for three processes, inverse beta decay for $\nu_1$ and $\nu_2$ and a direct beta decay for $\bar{\nu}_2$. The condensate structure of the flavor states is such that the absorption of a $\nu_2$ and the emission of a $\bar{\nu}_2$ are combined to yield the $\nu_e$ inverse beta decay amplitude.

\end{itemize}

From the formula \eqref{appendix2}, having for the Mandelstam variables, $s \simeq m_n^2, t \simeq (m_e - m_{\nu})^2 + 2 p_e p_{\nu} \cos \theta$, we can write the differential cross section as

\begin{eqnarray}\label{CrossSection1}
\nonumber \frac{d \sigma}{d \cos \theta} &=& \frac{1}{32 \pi} \frac{1}{m_n^2}\frac{p_e}{p_{\nu}} \left[ |M_{\nu_1}|^2 (E_{\nu,1}) + |M_{\nu_2}|^2 ( |E_{\nu,2}|) + |M_{\bar{\nu}_2}|^2  (- |E_{\nu,2}|) \right] \\ \nonumber
&=& \frac{G_F^2}{4 \pi}|V_{ud}|^2 \frac{m_p E_e p_e}{m_n} \Bigg[\left( A_1(s_{\nu}) (f^2 + 3g^2) + B_1(s_{\nu}) (f^2-g^2) v_e \cos \theta  \right)\frac{\cos^2 \Theta \cos^2 \bar{\Theta}}{v_{\nu,1}}  \\
\nonumber &+& \left( A_2(s_{\nu}) (f^2 + 3g^2) + B_2(s_{\nu}) (f^2-g^2) v_e \cos \theta  \right)\frac{\sin^2 \Theta \sin^2 \bar{\Theta} |U_{\pmb{p}_{\nu}}|^2}{v_{\nu,2}} \\
 &-& \left( A_{\bar{2}}(s_{\nu}) (f^2 + 3g^2) + B_{\bar{2}}(s_{\nu}) (f^2-g^2) v_e \cos \theta  \right)\frac{\sin^2 \Theta \sin^2 \bar{\Theta} |V_{\pmb{p}_{\nu}}|^2}{v_{\nu,2}} \Bigg] \ .
\end{eqnarray}

Here we have explictly written the neutrino energy argument in order to keep track of the change of sign in the last term. The functions $A$ and $B$ are defined as
\begin{eqnarray}
A_{j}(s_{\nu}) &=& 1 - 2 s_{\nu} v_{\nu,j},
 \\
 B_{j}(s_{\nu})  &=& v_{\nu,j} - 2s_{\nu},
  \\
  A_{\bar{j}}(s_{\nu}) &=& 1 + 2 s_{\nu} v_{\nu,j}, 
  \\
  B_{\bar{j}}(s_{\nu}) &=& -v_{\nu,j} - 2s_{\nu},
 \end{eqnarray}  
   in terms of the (positive) neutrino velocities $v_{\nu,j} = \frac{p_{\nu}}{|E_{\nu,j}|}$. Notice the change of sign in $A_{\bar{j}}, B_{\bar{j}}$ prompted by the substitution $|E_{\nu}| \rightarrow - |E_{\nu}|$.
The integral over $\cos \theta$ is trivial and cancels all the $B$ terms.
For the nonrelativistic neutrinos considered here $v_{\nu,j} = \frac{p_{\nu,j}}{E_{\nu,j}} \ll 1$, so that $A_{j} (s_{\nu}) = A_{\bar{j}} (s_{\nu}) \rightarrow 1$ and we can neglect the difference $v_{\nu,1} - v_{\nu,2} = \frac{p_{\nu}(m_2 -m_1)}{m_1 m_2} \ll 1$. We then multiply Eq. \eqref{CrossSection1} by, $v_{\nu,1} \simeq v_{\nu,2}$, and the Fermi function, $F(Z,E_e) = \frac{2 \pi \eta}{1-e^{-2\pi \eta}}$, with $\eta = \alpha \frac{E_e}{p_e}$, to obtain the capture cross section

\begin{equation}\label{CrossSection2}
 \sigma = \frac{G_F^2}{2 \pi} |V_{ud}|^2 F(Z,E) \frac{m_p E_e p_e}{m_n} (f^2 + 3g^2) \left[\cos^2 \Theta \cos^2 \bar{\Theta} + \sin^2 \Theta \sin^2 \bar{\Theta} \left(|U_{\pmb{p}_{\nu}}|^2 - |V_{\pmb{p}_{\nu}}|^2 \right) \right] \ .
\end{equation}

\begin{figure}[h]
 \includegraphics[width=0.5\linewidth]{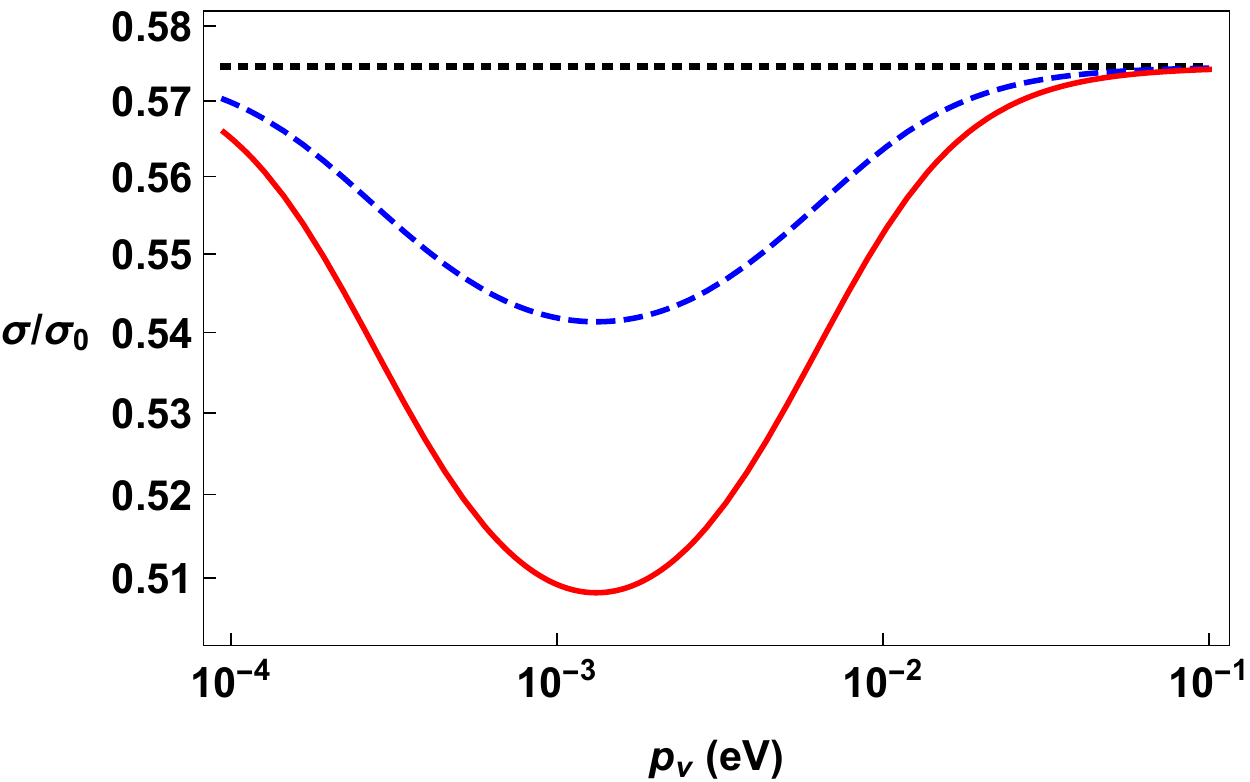}
 \caption{(color online): Plot of the ratio $\frac{\sigma (p_{\nu})}{\sigma_0}$ as a function of the neutrino momentum $p_{\nu}$ in the range $[10^{-4},10^{-1}] \  \mathrm{eV}$. The black dotted line corresponds to the Pontecorvo cross section (Eq. \eqref{CrossSection4}) $\sigma (p_{\nu}) = \sigma_P$; the blue dashed line is the Pontecorvo-Dirac cross section (Eq. \eqref{PDCrossSection} $\sigma (p_{\nu}) = \sigma_{PD}$, and the red solid line is the flavor Fock space cross section (Eq. \eqref{CrossSection5}) $\sigma (p_{\nu}) = \sigma_{F}$. The ratio of the cross section for decoupled neutrinos $\frac{\sigma_0}{\sigma_0}$ is $1$ by definition, and has not been plotted. The parameters have been chosen as $\sin^2 \Theta = \sin^2 \Theta_{12} = 0.307 $, $\Delta m^2 = \Delta m^2_{12} = 7.53 \times 10^{-5} \mathrm{eV}^2$ and $m_1 = 2 \times 10^{-4} \mathrm{eV}$.}
\end{figure}

The cross section for the various mixing schemes summarized in Table I is:

\begin{enumerate}
 \item \emph{Decoupled Pontecorvo}

 Unsurprisingly, the cross section in this case matches the result of Ref. \cite{Long2014}

\begin{equation}\label{CrossSection3}
 \sigma_0 = \frac{G_F^2}{2 \pi} |V_{ud}|^2 F(Z,E) \frac{m_p E_e p_e}{m_n} (f^2 + 3g^2) \ ,
 \end{equation}

 showing that Eq. \eqref{CrossSection2} constitutes a proper generalization.

 \item \emph{Pontecorvo}

 Here $U \rightarrow 1, V \rightarrow 0$ and $\bar{\Theta} \rightarrow \Theta$ and the cross section reads

 \begin{equation}\label{CrossSection4}
 \sigma_P = \frac{G_F^2}{2 \pi} |V_{ud}|^2 F(Z,E) \frac{m_p E_e p_e}{m_n} (f^2 + 3g^2) \left[\cos^4 \Theta + \sin^4 \Theta \right] \ .
 \end{equation}
 We can see that the difference with respect to the first case is only in the multiplicative factor, $\cos^4 \Theta + \sin^4 \Theta$.

 \item \emph{Pontecorvo-Dirac}

 The cross section has a non-trivial dependence on the neutrino momentum via $|U_{\pmb{p}_{\nu}}|$:

 \begin{equation}\label{PDCrossSection}
 \sigma_{PD} = \frac{G_F^2}{2 \pi} |V_{ud}|^2 F(Z,E) \frac{m_p E_e p_e}{m_n} (f^2 + 3g^2) \left[\cos^4 \Theta  + \sin^4 \Theta  |U_{\pmb{p}_{\nu}}|^2  \right] \ .
\end{equation}

 Recall that the same result is obtained within the flavor Fock space approach if the negative energy term is discarded.

 \item \emph{Flavor Fock space states}

 For the cross section we have

 \begin{equation}\label{CrossSection5}
  \sigma_F =  \frac{G_F^2}{2 \pi} |V_{ud}|^2 F(Z,E) \frac{m_p E_e p_e}{m_n} (f^2 + 3g^2) \left[\cos^4 \Theta  + \sin^4 \Theta  \left(|U_{\pmb{p}_{\nu}}|^2 - |V_{\pmb{p}_{\nu}}|^2 \right) \right] \ .
 \end{equation}

 This result is acceptable  if the negative energy term is interpreted as a condensate hole term, otherwise one gets Eq. \eqref{PDCrossSection}.

\end{enumerate}

The capture cross sections of Eqs. \eqref{CrossSection3}, \eqref{CrossSection4}, \eqref{PDCrossSection} and \eqref{CrossSection5} refer to the neutrino capture on a free neutron. In order to obtain the capture cross sections on tritium $\nu_e + \ce{^{3}H} \rightarrow \ce{^{3}He} + e^{-}$, all these expressions have to be changed according to the procedure described in the ref. \cite{Long2014}. In particular, the nucleon masses $m_n , m_p$ are replaced with the masses of the atomic species $m_{\ce{^{3}H}}$ and $m_{\ce{^{3}He}}$, the form factors $f^2, 3g^2$ are replaced with the nuclear matrix elements $\langle f_F\rangle^2, (g_A/g_V)^2 \langle g_{GT}\rangle^2$, and the electron kinetic energy $E_e - m_e$ must be modified accordingly. Let us set, for convenience
\begin{eqnarray*}
 \sigma_{n0} &=&  \frac{G_F^2}{2 \pi} |V_{ud}|^2 F(Z,E) \frac{m_p E_e p_e}{m_n} (f^2 + 3g^2) \\
 \sigma_{T0} &=&  \frac{G_F^2}{2 \pi} |V_{ud}|^2 F(Z,E) \frac{m_{\ce{^{3}He}} E_e p_e}{m_{\ce{^{3}H}}} (\langle f_F\rangle^2 + (g_A/g_V)^2 \langle g_{GT}\rangle^2)
\end{eqnarray*}
denoting, respectively, the capture cross sections for decoupled neutrinos on free neutrons $\sigma_{n0}$ (Eq. \eqref{CrossSection3}) and on tritium $\sigma_{T0}$. Notice that the ratios between the cross section in a given scheme $\sigma =$ $\sigma_{P}$ (Pontecorvo), $\sigma_{PD}$ (Pontecorvo-Dirac), $\sigma_F$ (Flavor Fock space), and the reference cross section for decoupled neutrinos $\sigma_0$ are the same for the capture on free neutrons and on tritium, i.e. $\frac{\sigma_{n \alpha}}{\sigma_{n 0}} = \frac{\sigma_{T \alpha}}{\sigma_{T 0}}$ for each $\alpha =$ P, PD, F.
The ratios are plotted in Fig. 1  for neutrino momenta $p_{\nu}$ in the range $[10^{-4},10^{-1}] \  \mathrm{eV}$.

\begin{figure}[h]
 \includegraphics[width=0.7\linewidth]{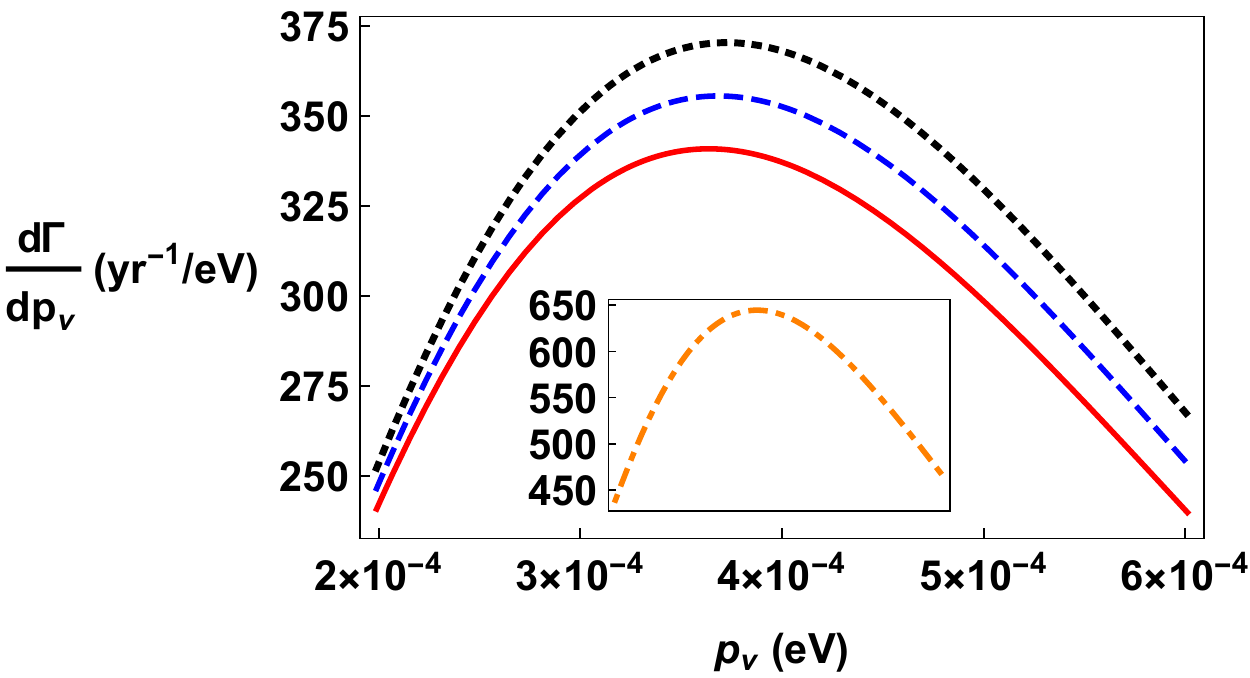}
 \caption{(color online): Plot of the differential capture rate (Eq. \eqref{capturerate}) $\frac{d \Gamma (p_{\nu})}{d p_{\nu}}$ as a function of the neutrino momentum $p_{\nu}$ in the range $[2\times 10^{-4},6 \times 10^{-4}] \  \mathrm{eV}$. The differential rate is of course adimensional, but is expressed in terms of $\mathrm{yr}^{-1}/\mathrm{eV}$ for ease of comparison with the previous results.
 The black dotted line corresponds to the Pontecorvo cross section (Eq. \eqref{CrossSection4}) $\sigma (p_{\nu}) = \sigma_P$, the blue dashed line is the Pontecorvo-Dirac cross section (Eq. \eqref{PDCrossSection} $\sigma (p_{\nu}) = \sigma_{PD}$ and the red solid line is the Flavor Fock space cross section (Eq. \eqref{CrossSection5}) $\sigma (p_{\nu}) = \sigma_{F}$. In the inset we plot the capture rate (orange dotdashed line) corresponding to the decoupled Pontecorvo cross section (Eq. \eqref{CrossSection3}). We have assumed a sample of tritium with total mass $M_S = 100 \  \mathrm{g}$. The other parameters have been chosen as in figure 1.}
\end{figure}

The capture rate on a sample of tritium of mass $M_S$ can be expressed in terms of the capture cross section $\sigma$ as

\begin{equation}
 d \Gamma = \sigma N_T d n_{\nu} \ ,
\end{equation}
where $N_T = \frac{M_S}{m_{\ce{^{3}He}}}$ is the total number of tritium nuclei in the sample and $dn_{\nu}$ is the (differential) number density of neutrinos per degree of freedom. Within the sudden freeze-out approximation, the phase space distribution of neutrinos is the redshifted distribution function that was realized at the decoupling epoch. At redshift $z$, the number density reads
\begin{equation}
 dn_{\nu} (z) = \frac{d^3 p(z)}{(2\pi)^3}\frac{1}{e^{\frac{p(z)}{T_{\nu (z)}}}+1} = \frac{p^2 (z) d p(z)}{2 \pi^2}\frac{1}{e^{\frac{p(z)}{T_{\nu (z)}}}+1}
\end{equation}
where we have performed the solid angle integration in the last step. Here, $p(z)= \frac{1+z}{1 + z_{FO}}p_{FO}$, $T_{\nu}(z)= \frac{1+z}{1 + z_{FO}}T_{\nu,FO}$ and $z_{FO} = 6 \times 10^{10}$, \cite{Long2014} with the label $FO$ denoting the quantities at the freeze-out. At the current epoch $z=0$ we have
\begin{equation}
 dn_{\nu} (z=0) = dp_{\nu} \frac{p^2_{\nu}}{2 \pi^2} \frac{1}{e^{\frac{p_{\nu}}{T_{\nu}}}+1}
\end{equation}
where $p_{\nu} = p(z=0)$ and $T_{\nu} = T_{\nu} (z = 0) \simeq 0.168 \times 10^{-3} \mathrm{eV}$. Finally, the differential capture rate becomes
\begin{equation}\label{capturerate}
 \frac{d\Gamma}{dp_{\nu}} = N_T \frac{\sigma (p_{\nu}) p_{\nu}^2}{2 \pi ^2}\frac{1}{e^{\frac{p_{\nu}}{T_{\nu}}}+1}
\end{equation}
with the capture cross section $\sigma(p_{\nu})$. The capture rate of eq. \eqref{capturerate} is twice as large for Majorana neutrinos \cite{Long2014}. In Fig.(2), we plot the differential capture rate for the various schemes discussed above.

As it is evident from Fig.(2), the capture rate for coupled neutrinos is smaller than the capture rate for decoupled neutrinos due to the factor $\cos^4 \Theta + \sin^4 \Theta < 1$. The dependence on the neutrino momentum is different for all the four cases considered, due to the presence (or the absence) of the Bogoliubov coefficients $U$ and $V$. In particular, the capture rate for neutrinos is the lowest for the flavor Fock space states (the red line in Fig.(2)). We remark that it is this last case that corresponds to the condensed flavor vacuum, and hence to a possible dark matter component. As anticipated, the difference between the various approaches is mostly relevant for neutrino momenta comparable to the neutrino masses $p_{\nu} \simeq m_{\nu}$. For larger momenta all the three curves of Fig.(2) converge to the Pontecorvo result (the black dotted line). A similar conclusion holds for the capture cross sections, as it can be seen from Fig.(1).

\section{Conclusions}

We have analyzed the possibility to test the QFT of neutrino mixing and the predicted contribution of the flavor vacuum to the dark matter by means of low energy neutrino capture experiments. We have considered the possible schemes of neutrino mixing and the related flavor states definition in the computation of the capture rate of neutrino absorption by tritium. We have shown that the capture cross section and the capture rate depend on the definition of the flavor states and are significantly different for very small neutrino momenta $p_{\nu} \simeq m_{\nu}$.  This amounts to an observable trace of the Bogoliubov coefficients that characterize the condensation of the flavor vacuum in the flavor Fock space model.
Therefore, experiments such as PTOLEMY, projected to reveal  the cosmic neutrino background (neutrinos with momenta $p_{\nu} \simeq m_{\nu}$),  can not only distinguish among the possible mixing schemes, but also provide an indirect evidence of the flavor condensate and of the dark matter contribution induced by the neutrino mixing.
An analogous mechanism of flavor condensation is also predicted for mixed bosons \cite{BosonMixing} and the possible role of the boson flavor vacuum as a dark energy component has been investigated \cite{FDM1}. Future studies 
on the pure annihilation type radiative $B$ meson decays $B^0 \rightarrow \phi \gamma$ and $B_s \rightarrow \rho^{0}(\omega)\gamma$ \cite{MesonMixing1} and on $\omega-\phi$ meson mixing \cite{MesonMixing2}, could allow the detection of observable signatures of the boson flavor vacuum condensation in meson mixing phenomena. Due to the larger mass differences involved with respect to neutrino mixing, it is reasonable to prospect even more significant observable effects on meson mixing, due to the structure of the flavor vacuum.

\section*{Acknowledgements}
Partial financial support from MIUR and INFN is acknowledged.
A.C. also acknowledges the COST Action CA1511 Cosmology
and Astrophysics Network for Theoretical Advances and Training Actions (CANTATA).

\appendix

\section{Cross section for generalized amplitude}

In this appendix we work out the cross section for amplitudes containing more than one energy delta function. Following Weinberg \cite{Weinberg} the differential probability for the $I \rightarrow F$ transition is
\begin{equation}
 dP(I \rightarrow F) = \left[ \frac{(2 \pi)^3}{V} \right]^{N_I} |S_{FI}|^2 dF
\end{equation}
where $N_I$ is the number of particles in the state $\ket{I}$, $dF$ is the phase space volume element for the final state and we are assuming that the system is confined in a box of volume $V$. Likewise, we assume that the interaction is switched on for a finite time $T$. Suppose that the $S$- matrix has the form
\begin{equation}\label{appendix1}
 S_{FI} = - 2 \pi i \delta^{3}_{V}(\pmb{P}_F - \pmb{P}_i) \sum_{j} \delta_T (E_{I,j}-E_{F}) M_{FI; j} \ .
\end{equation}
Here $\delta_V^3 (\pmb{p}) = \frac{1}{(2\pi)^3} \int_V d^3 x e^{i \pmb{p}\cdot \pmb{x}}$ and $\delta_T (E) = \frac{1}{2 \pi} \int_{-\frac{T}{2}}^{\frac{T}{2}} dt e^{-i E t}$ are the finite volume and the finite time delta functions. The matrix elements $M_{FI;j}$ refer to the processes in which the initial state has total energy $E_{I,j}$, total momentum $\pmb{P}_I$ and the final state has total energy $E_{F}$ and total momentum $\pmb{P}_F$. Upon squaring the $S$-matrix element of eq. \eqref{appendix1} we shall get mixed terms proportional to distinct delta functions $\propto \delta_T(E_{I,j}-E_{F}) \delta_T(E_{I,j'}-E_{F})$. These terms impose incompatible conditions on the final energy ($E_F = E_{I,j} = E_{I,j'}$, whereas $E_{I,j} \neq E_{I,j'}$ by hypothesis) and therefore yield a zero contribution. The only surviving terms are those with the same energy delta function:
\begin{equation}
 |S_{FI}|^2 = (2\pi)^2 \left[\delta^{3}_{V}(\pmb{P}_F - \pmb{P}_i) \right]^2 \sum_{j} \left[\delta_T (E_{I,j}-E_{F}) \right]^2 |M_{FI;j}|^2 \ .
\end{equation}
The squared delta functions can be easily interpreted within the finite volume and finite time hypothesis \cite{Weinberg}, giving for the differential probability
\begin{equation}
 dP(I \rightarrow F) = (2 \pi)^2 \left[ \frac{(2 \pi)^3}{V} \right]^{N_I -1} \frac{T}{2 \pi} \delta^{3}_{V}(\pmb{P}_F - \pmb{P}_i) \sum_{j}\delta_T (E_{I,j}-E_{F})|M_{FI;j}|^2 \ .
\end{equation}
Specializing to $N_I = 2$ and taking the infinite volume/infinite time limits the differential cross section is
\begin{equation}\label{appendix2}
 d \sigma (I \rightarrow F) = \frac{dP(I\rightarrow F)}{u_I} \frac{V}{T} = (2\pi)^4 u^{-1}_{I} \sum_{j} \delta^4 (P_{I;j}^{\mu}- P_{F}^{\mu}) |M_{FI;j}|^2 dF
\end{equation}
where $u_I$ is the relative velocity.

\end{document}